\documentclass[%
 preprint, 
linenumbers,
 amsmath,amssymb,
 aps, physrev
]{revtex4-2}

\usepackage{graphicx}% Include figure files
\usepackage{dcolumn}% Align table columns on decimal point
\usepackage{bm}% bold math
\usepackage[T1]{fontenc}% Font encoding for Polish characters
\usepackage{braket} % ket|> and bra <|
\usepackage{epstopdf}

\begin{document}
\nolinenumbers
\preprint{APS/123-QED}

\title{Efficient Storage of Multidimensional Telecom Photons in a Solid-State Quantum Memory}

\author{Zongfeng Li}
\author{Yisheng Lei}%
\author{Trevor Kling}%
\affiliation{%
 Department of Electrical and Computer Engineering and Applied Physics Program, Northwestern University, Evanston, IL 60208, USA
}

\author{Mahdi Hosseini}%
\email{Contact author: mh@northwestern.edu}
\affiliation{%
 Department of Electrical and Computer Engineering and Applied Physics Program, Northwestern University, Evanston, IL 60208, USA
}%
\affiliation{%
 Elmore Family School of Electrical and Computer Engineering, Purdue University, West Lafayette, Indiana 47907, USA
}%
\affiliation{%
 Department of Physics and Astronomy, Purdue University, West Lafayette, Indiana 47907, USA
 }%

\date{\today}% It is always \today, today,

\begin{abstract}

Efficient storage of telecom-band quantum optical information represents a crucial milestone for establishing distributed quantum optical networks. Erbium ions in crystalline hosts provide a promising platform for telecom quantum memories; however, their practical applications have been hindered by demanding operational conditions, such as ultra-high magnetic fields and ultra-low temperatures. In this work, we demonstrate the storage of telecom photonic qubits encoded in polarization, frequency, and time-bin bases. Using the atomic frequency comb protocol in an Er$^{3+}$-doped crystal, we developed a memory initialization scheme that improves storage efficiency by over an order of magnitude under practical experimental conditions. Quantum process tomography further confirms the memory's performance, achieving a fidelity exceeding 92\%.

\end{abstract}

\maketitle

%\tableofcontents

\section{Introduction}

Optical quantum networks have numerous proposed applications such as distributed quantum computing, blind quantum computing, secure communications, and distributed quantum sensing \cite{kimble2008quantum, wehner2018quantum}. Due to the exponential optical loss in optical fibers and the fundamental inability of quantum information amplification, quantum memory devices are sought to relay and synchronize quantum information distributed over long distances  \cite{briegel1998quantum, duan2001long, lei2023quantum} . To enable long-range networking, various quantum repeater schemes have been proposed \cite{duan2001long, muralidharan2016optimal,awschalom2021development, azuma2023quantum}. Quantum repeaters based on absorptive quantum memories have the potential advantages of long-life storage \cite{ma2021one}, multiplexed operations \cite{simon2007quantum, yang2018multiplexed, gu2024hybrid}, photonic integration \cite{zhou2023photonic, perminov2023integrated}, and interfacing with ubiquitous entangling sources \cite{lago2021telecom, liu2021heralded, thomas2024deterministic}. In addition, quantum memories have other applications including, but no limited to, single photon generation \cite{kaneda2019high}, multi-photon synchronization \cite{kaneda2017quantum, chrapkiewicz2017high}, linear optical quantum computing \cite{knill2001scheme}, quantum sensing \cite{zaiser2016enhancing, ding2020quantum}, and fundamental science in space \cite{mol2023quantum}.

Quantum memories with rare-earth-ions in solids are under active developments \cite{guo2023rare, zhou2023photonic, lei2023quantum}. Elementary quantum networks have been demonstrated with rare-earth ion ensembles in crystals interfaced with entangling photon sources and wavelength conversion to accommodate memory wavelength, which is mostly outside the telecom band \cite{liu2021heralded, lago2021telecom}. Typically, atomic frequency comb (AFC) protocol  \cite{afzelius2009multimode} is employed for storage. Erbium ions, Er$^{\text{3+}}$, in solids have optical transitions in the telecom C-band, an unparalleled property that is suitable for integration with existing optical fiber networks. Over 1~s coherence time of nuclear spin in a high magnetic field and milliseconds in zero field has been reported with Er$^{\text{3+}}$: YSO crystal \cite{ranvcic2018coherence,rakonjac2020long}, and memory-source integration has been demonstrated as well\cite{jiang2023quantum}. Using Er$^{\text{3+}}$:CaWO$_{\text{4}}$ crystals, an electronic spin coherence time of 23~ms has been shown \cite{le2021twenty}, and single spin-photon entanglement has been demonstrated \cite{uysal2024spin}. The practical deployment of Er-doped solids as quantum memories enables advancements beyond passive memory elements \cite{arnold2024all}, significantly extending communication distances, while offering a platform for deterministic entanglement and coherent control of information.

The recent development in Er crystals have relied on dilution refrigeration or high magnetic fields (about 7 Tesla) to achieve certain memory milestones. The presence of a half-integer spin for the Er$^{3+}$ demands both low temperatures and a high magnetic field to freeze the coherence-degrading spin-flip transitions \cite{ranvcic2018coherence}.  Additionally, the low branching ratio associated with the employed shelving transitions and the presence of large number of hyperfine states complicate the pumping of spectral features in erbium \cite{hastings2008zeeman} compared to other ion candidates. For these reasons, the initialization and operation of Er-based quantum memories has proved difficult posing limitations on its practical deployment when it comes to multiplexed processing. 

A storage efficiency of up to 22\% has been observed in an $^{167}$Er:YSO crystal with a 7 T field at 1.8 K \cite{stuart2021initialization}.  
Other approaches have utilized a moderate magnetic field in a dilution refrigerator to achieve the same spin-freezing, but with a much lower storage efficiency of 0.2\% \cite{craiciu2019nanophotonic}. Multimode quantum storage with erbium doped fibers at 10 mK (dilution refrigerator) have been demonstrated \cite{wei2024quantum}, however it suffers from low storage efficiency and short storage time. 

Despite significant advancements in telecom quantum memory development, achieving a scalable solution for practical quantum optical networks requires overcoming the reliance on ultra-high magnetic fields and dilution refrigerators. Additionally, efficient multidimensional qubit storage—essential for the realization of multiplexed quantum networks—has yet to be demonstrated in the telecom band.

In this article, we report the development of an efficient and multidimensional quantum storage in an AFC quantum memory based on an Er$^{\text{3+}}$:YSO crystal, without using superconducting magnets nor a dilution refrigerator. We develop a simple but efficient spectral hole pumping method to perform the spectral tailing required for the AFC storage in the telecom band. We demonstrate the storage and retrieval of single-photons in polarization, frequency,  and time-bin basis.  
Memory fidelity was evaluated using quantum process tomography for polarization qubits and intensity noise estimation for frequency and time-bin qubits.

\section{Experimental Platform}

We investigate an isotopically-purified erbium-167 doped in yttrium orthosilicate crystal ($^{167}$Er$^{3+}$:YSO) from Scientific Materials.  
The crystal dimensions are 2~mm~$\times$~3~mm~$\times$~4~mm, with the b-axis aligned with the 4~mm edge. Both D$_1$ and D$_2$ axes are in the 2~mm~$\times$~3~mm plane, at a 45$^\circ$ angle to the crystal edges.
Each Er$^{3+}$ ion substitutes for a Y$^{3+}$ ion in one of two sites with $C_1$ rotational symmetry.  Each of these two distinct sites are then divided into four orientation subsites, corresponding to the $C_2$ rotation and inversion symmetry.  While in general these subsites are magnetically inequivalent, by aligning the magnetic field in the D$_1$-D$_2$ plane of the crystal the magnetic equivalence can be restored \cite{sun2008magnetic}.

consist of 8 and  $^{4}$I$_{15/2}$6 Krathe mers doublets, resp $^{4}$I$_{13/2}$ectively, and then7. These doublets become
Applying a strong magnetic field allows the electron spin to be frozen\cite{ranvcic2018coherence}, reducing the state occupation to the eight hyperfine nuclear spin states. 
The lifetime of these states at low temperatures and high magnetic fields can be on the order of minutes \cite{ranvcic2018coherence}, which makes them suitable as shelving levels for a quantum memory. Additionally, defects in the host YSO materials produce fluctuations in the local crystal field, leading to inhomogeneous broadening of the ensemble transitions  For a dopant concentration of 0.005\%, this material has an exceptionally low inhomogeneous linewidth of 390 MHz, and can reach a narrow homogeneous linewidth of 73 Hz \cite{bottger2006spectroscopy}.

\begin{figure*}
    \centering
    \includegraphics[width=0.7\linewidth]{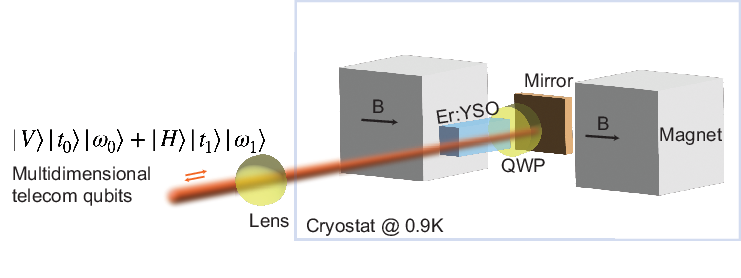}
    \caption{
        Schematic of the experimental system.  A $^{167}$Er$^{3+}$:YSO crystal is situated between two monolithic rare-earth magnets and placed within a 0.9 K cryostat.  The erbium atoms are addressed via a telecom-wavelength laser propagating along the $b$ axis of the crystal, passing through the crystal twice with a $\pi/2$ phase shift due to the mirror-quarter wave plate pair. The memory is supplied with frequency-, time-, and polarization-bin qubits, storing the information for a fixed period of time before re-emission.
    }
    \label{fig:optical_setup}
\end{figure*}

The crystal is cooled to 0.9 K and subjected to about 1.1 T of magnetic field from a set of parallel neodymium block magnets. A schematic of the experimental setup is shown in Fig.\ref{fig:optical_setup}. The laser wavelength at 1536.3 nm coincides with the $^{4}$I$_{15/2}$ $\leftrightarrow$ $^{4}$I$_{13/2}$ transition in Er$^{3+}$ for the first substitutional site of YSO.  The magnetic field is oriented at a $135^{\circ}$ angle from the D$_1$ axis of the crystal in the D$_1$-D$_2$ plane, while the optical field propagates along the $b$ axis.

\section{Results}

\subsection{Memory initialization} 

Compared to other rare-earth ions, the 10~ms-long excited state lifetime of Er$^{3+}$ ions significantly reduce the spectral tailoring performance. This is particularly a limitation when considering the branching ratio of less than 10\% \cite{hastings2008zeeman} and limited ground state lifetime imposed by the coupling between the electron spin to the lattice and other spins. This results in difficulties creating spectral holes and efficiently tailoring the absorption spectrum and initializing the memory \cite{lauritzen2008state}. 

In Er:YSO, the purpose of spectral tailoring is to shelve ions from a specific energy level to other hyperfine levels, thereby creating spectral regions with no absorption or weak absorption (i.e. spectral hole). The magnetic field can be aligned along a particular direction of the crystal, to have site 1 ions with largest gyromagnetic ratio \cite{sun2008magnetic}. The resulted level splitting and the low phonon occupation number at 0.9 K, enables extending the hole lifetime to more than 3 seconds (see S.I. Sec.E). We also observed an optical coherence time of $T_2=169~\mu s$ (see Fig.S1(c)) at 1.1 T of field measured using the two-pulse echo method \cite{macfarlane1997measurement}. 

To improve the efficiency of memory initialization, we devised an "interleaved pumping" scheme extending the efficiency by more than an order of magnitude. In traditional pumping protocols, continuous or semi-continuous pumping were employed for a duration much longer than the excited state lifetime. This approach works best for ions with a much shorter excited state lifetime, or when a weak pump is used and the hole lifetime is extended to minute scales (using $\sim$7~T of magnetic fields). We have observed that in the case of intermediate hole lifetimes, continuous pumping leads to inefficient initialization as the excited atoms undergo stimulated emission and return to the initial ground state. In the interleaved pumping scheme, the pump is switched off periodically for about 10~ms (in-loop delay) allowing atoms to decay with a higher probability to the shelved state. The pumping sequence is shown in Fig.2(a). 
A single or multiple AFC windows were created by sweeping the frequency of pump over the corresponding frequency window. 
The pumping cycle consists of a set of complex hyperbolic secant pulses that exhibit a square frequency spectrum, followed by an in-loop delay. \cite{minavr2010spin, jobez2016towards}.
Multiple pumping parameters were scanned to search for maximum storage efficiency (see S.I. Sec. B). 
Fig.2(b) shows the depth of the spectral hole, indicating the efficiency of the hole burning for different in-loop delays, as an example. 
At the end of the pumping cycle, an out-loop delay is applied to allow excited atoms to decay prior to the input qubits.
As shown in Fig.2(c), the spontaneous emission noise is negligible after 50~ms. The storage efficiency up to 6.1\% and 5.0\% for a classical and single photon pulse, respectively, is observed, with an out-loop delay of 100~ms. This is an enhancement of more an order of magnitude compared to previous work where low magnetic field was used \cite{craiciu2021multifunctional}.

\begin{figure}[!h]
    \centering
    \includegraphics[width=0.8\columnwidth]{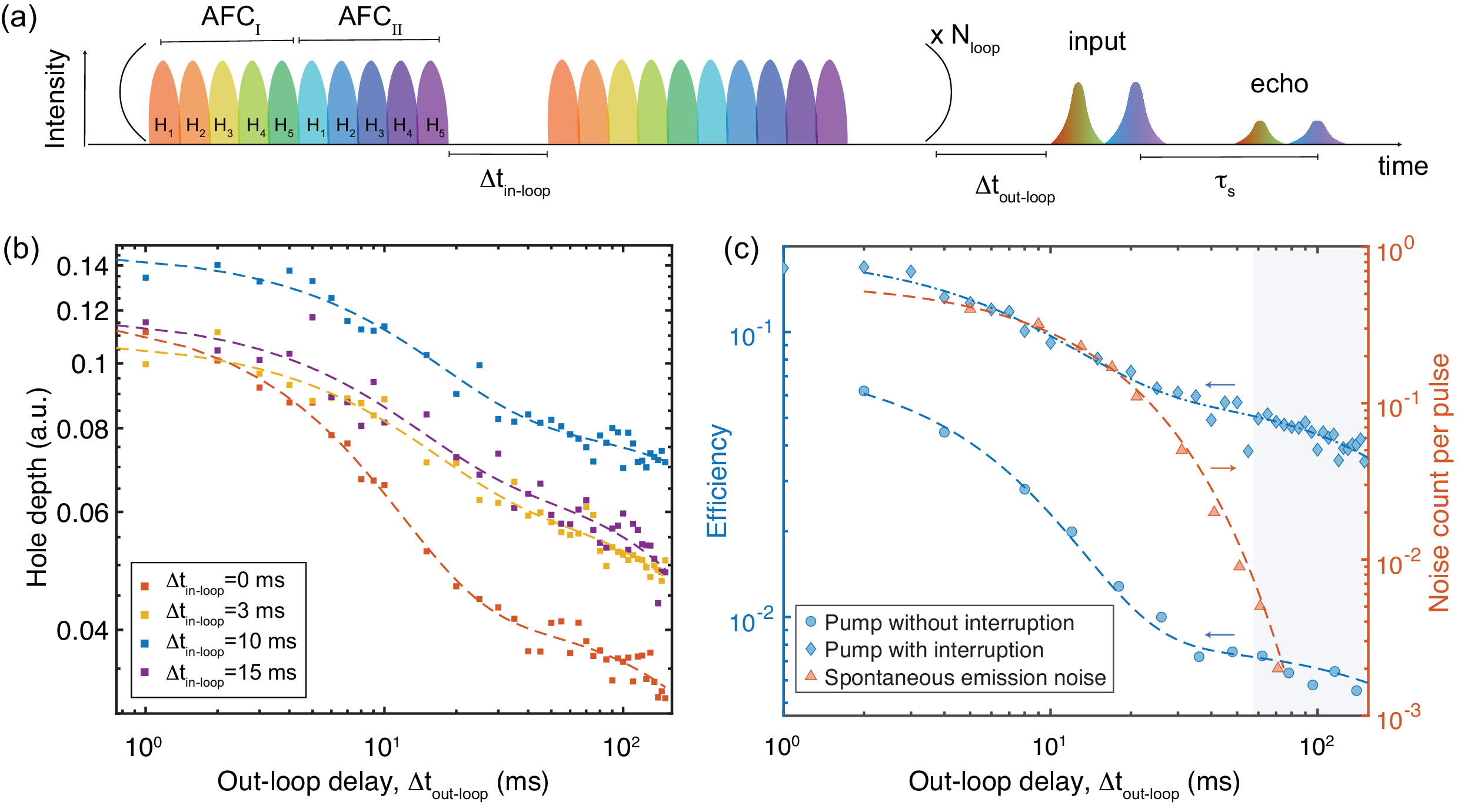}
    \caption{
    (a) Schematic of the interleaved optical pumping sequence for spectral tailoring of the atomic spectrum. The system includes five parameters to be optimized; the power, duration and spectral width of each hole $\mathrm{H}_i$, the number of batches performed $N_{loop}$, and the delay time between each batch of pulses $\Delta t_{in-loop}$.  After preparation, the system simultaneously stores two qubits encoded on multiple degrees of freedom for a time $\tau_{s}$. 
    (b) By varying the pumping sequence parameters, the depth of the spectral hole is modulated.  Choosing an optimal set of pumping parameters maximizes the spectral tailoring, which improves storage efficiency. (c) Blue: The AFC memory efficiency with interleaved (diamond) or continuous (circle) pumping vs. waiting time, $\Delta t_{out-loop}$, after pump. Orange: Average noise photon counting due to the spontaneous emission from the excited state. For $\Delta t_{out-loop}>50$ms the spontaneous emission noise is found to be negligible (shaded region). Dashed lines in (b) and (c) are double-exponential fits.
    }
    \label{fig:pumpingSequence}
\end{figure}

\begin{figure}[!h]
    \centering
    \includegraphics[width=\columnwidth]{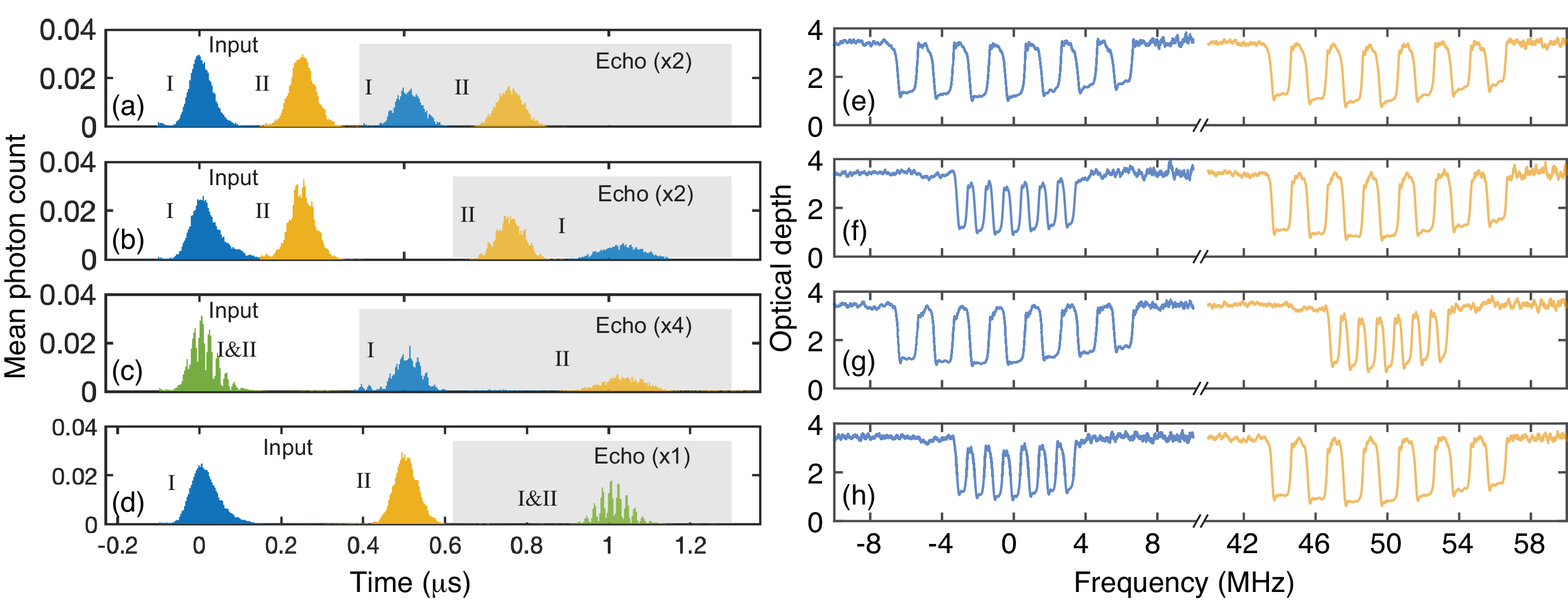}
    \caption{
    Memory performance at the single-photon level. Coherent control over the storage of photons with two degrees of freedom; frequency and time is shown. By burning spectrally-separated AFC structures with varying parameters (right column), independent modulation of the storage time of each photon is achieved (left column). By storing qubits of varying frequency for different durations, we perform pulse processing, including (a,e) FIFO storage, (b,f) FILO storage, and (c-d, g-h) frequency-dependent delay. I and II refer to the first and second time-bin of input and echo light pulses. 
    }
    \label{fig:timefrequency}
\end{figure}

For atomic frequency combs with square comb profiles created by adiabatic pulses, the storage efficiency can be estimated using the following equation \cite{bonarota2010efficiency, jobez2016towards},
\begin{equation}\label{Eqa1}
\eta = \frac{d^2}{F^2}\text{exp}(-\frac{d}{F})\text{sinc}^2(\frac{\pi}{F})\text{exp}(-d_0), 
\end{equation}
where $d$ is the optical depth (OD) of the atomic medium before optical pumping, $F$ is the finesse of the combs, and $d_0$ is the OD corresponding to the background absorption after spectral tailoring.

Considering a fixed optical depth $d$, different pumping protocols can affect the background absorption and the AFC finesse. In our case, the OD is approximately $d\sim$3.3. The interleaved pumping returns a comb finesse of about $F\sim$3 and a background absorption OD of $d_0\sim$1.3. Theoretical estimated value for efficiency is $\sim$7.5\% that is in agreement with 6\% efficiency measured, given uncertainty in the OD estimation. We experimentally observed (Fig.2(c)) that the interleaved pumping results in about 8-fold enhancement of efficiency when compared to the traditional continuousness pumping.

\subsection{Multidimensional storage} 

To demonstrate the compatibility of multiple degrees of freedom (DOFs) of photonic qubits with the memory, we performed storage experiments across frequency, time, and polarization bases, as well as multidimensional storage involving combinations of these DOFs.
The frequency and time-bin storage is shown in Fig.\ref{fig:timefrequency}.
Two AFC windows of bandwidth 8 MHz \& 16 MHz separated by 50 MHz are prepared near the central peak of the absorption spectrum, with storage time between 0.5 $\mu$s or 1 $\mathrm{\mu}$s.
The input pulses at the two time-bins have different center frequencies matching that of two AFCs, each containing five spectral holes (comb lines). The delay can be designed for each AFC to retrieved the two time-bin data in either first-in-first-out (FIFO) or first-in-last-out (FILO) order. The frequency bins can be coherently separated (interfered) when they enter the memory at the same (different) time bins. This frequency-dependent delay can be used to coherently process qubits with frequency difference indistinguishable by gratings or other conventional frequency processors.  
The slightly broadened echos with 1 $\mu$s storage time in Fig.\ref{fig:timefrequency}(b-c) are due to the narrower corresponding AFC bandwidth. The imperfect input interference fringes in Fig.\ref{fig:timefrequency}(c) are caused by the Etalon leakage (0.9\%) of the 4 GHz EOM sideband.

\subsection{Process Tomography}

\begin{figure}[!h]
    \centering
    \includegraphics[width=0.7\columnwidth]{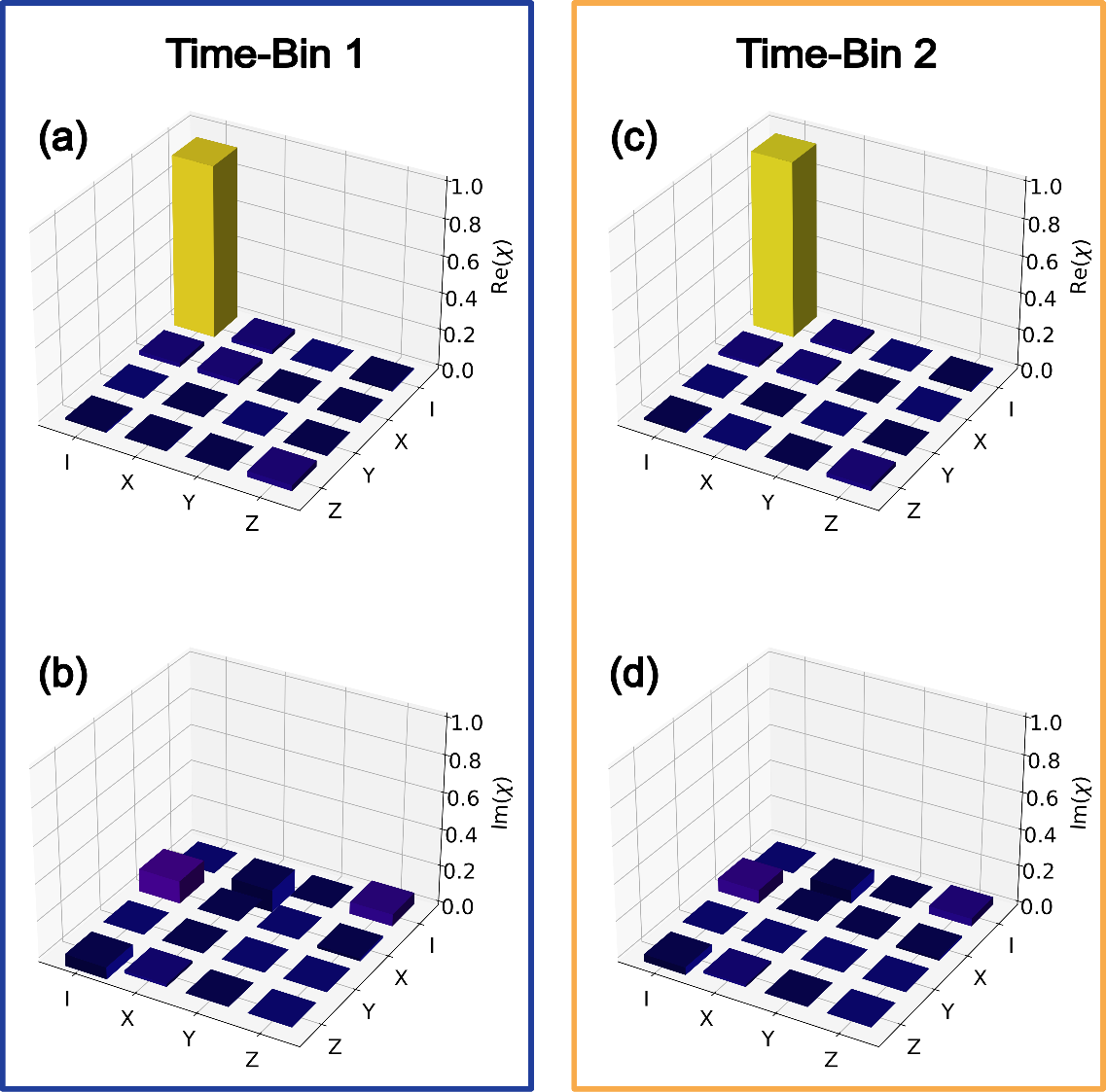}
    \caption{
    Real and imaginary components of the reconstructed process matrix for time-bin 1 (blue) and time-bin 2 (yellow) after a 0.5 $\mathrm{\mu}$s storage. The real components of the process matrices (a, c) are near unity for both echos, demonstrating high polarization stability during storage. The imaginary components of the process matrix contains nonzero elements, indicating the system acquires an uncompensated phase during the storage process.
    }
    \label{fig:polarizationChi}
\end{figure}

To evaluate the performance of the memory for storage of an arbitrary polarization state, we performed quantum process tomography of the quantum memory. The goal of the process tomography is to find the map that corresponds to an arbitrary, unknown process. Generally, this corresponds to determining some coefficient of matrix $\chi$ for a specified set of basis operators $\{A_n\}$, representing a map as shown in Equation \ref{eq:tomography}.

\begin{equation}\label{eq:tomography}
    \mathcal{E}(\rho) = \sum_{n,m} \chi_{nm} A_{n} \rho A^{\dagger}_{m}
\end{equation}

To determine $\chi$ from experimental data, we employed a standard minimum-value optimization procedure \cite{o2004quantum} for both a photodiode and a single-photon measurement. In the case of the photodiode measurement (see S.I. Fig.S5 (b)), input pulses are similar in amplitude to the pump field to produce the AFC.  For the single-photon operation, the input pulses are attenuated to fewer than one photons per shot.  We performed sixteen measurements for four orthogonal polarization states: $\{ |H\rangle, ~|V\rangle, ~|D\rangle = \frac{1}{\sqrt{2}}(|H\rangle + |V\rangle), ~|R\rangle = \frac{1}{\sqrt{2}}(|H\rangle - i|V\rangle) \}$. 
Three half waveplates was used to compensate the phase difference between 2 polarizations caused by the crystal ((see S.I. Sec. D).
To evaluate the accuracy of the memory process, we compute the process fidelity of the resulting map as $\mathcal{F} = \text{Tr}(\sqrt{\sqrt{\chi_{exp}}\chi_{ideal}\sqrt{\chi_{exp}}})^2$.  For a memory process the input state should remain unchanged, so the ideal $\chi$ matrix has a unit value for the identity term and zero elsewhere, corresponding to the map $\mathcal{E}_{ideal}(\rho) = I\rho I^\dagger$ where we have chosen an operator basis of Pauli operators $\{I, \sigma_x, -i\sigma_y, \sigma_z\}$ \cite{chuang1997prescription}, or a matrix with $\chi_{00} = 1$ and zeros elsewhere.

Tomography was completed for the simultaneous storage of two pulses at different time bins, separated in frequency by 50 MHz and stored for 0.5 $\mathrm{\mu}$s, the same procedure as is shown in Fig.\ref{fig:timefrequency}(a).  The tomography of the classical echos displayed a process fidelity of 94.9\% for the first pulse and 97.6\% for the second pulse.  Fig.\ref{fig:polarizationChi} shows the $\chi$ matrix elements for the single-photon tomography; in this case, the first echo demonstrated a process fidelity of 92.3\%, while the second echo had a slightly improved fidelity of 95.5\%.  The difference in fidelity of storage in these two cases is due to decreased interaction strength between the atoms and optical pulses of varying frequencies.  The largest non-identity component for the two pulses was the $ZZ$-component for the first echo, with a magnitude of $0.032$.  However, both maps had small but non-negligible complex components on the off-diagonal elements $IX$, $IZ$ (largest amplitude $0.123i$).  These components were matched with the conjugated values for the $XI$ and $ZI$ components, which suggests that the angled quarter-wave plate was insufficient to remove all phase effects.  In particular, the magnitude of these components varies between the first and second echo, with the first echo having a more pronounced rotation.

\section{Discussion}

There exist a number of avenues by which the performance of the system can yet be improved.  Adjusting the optical path through the cryostat can improve the polarization storage fidelity by removing dielectric-induced birefringence from the mirrors.  
By better aligning the magnetic field with respect to D$_1$ and D$_2$ axes or using site 2 ions of Er:YSO, a larger magnetic g-factor can be achieved \cite{sun2008magnetic}, leading to an enhanced ground state lifetime and spectral tailoring.(see S.I. Sec. E). 
Furthermore, the magnitude of the magnetic field may be improved by focusing the field through a set of wedges composed of a magnetic material.  Such elements could be incorporated directly into the mounting system for the crystals inside the cryostat.  
Besides, host crystals with low densities of nuclear spins such as CaWO$_\text{4}$ can be used to further increase the ground state lifetime and optical coherence time of erbium ions.

The spin-polarization initialization method \cite{stuart2021initialization} can be incorporated with the interleaved pumping scheme developed here to further enhance the memory performance \cite{baldit2010identification}.
With the ground state lifetime being extended to 10~s, more efficient spectral tailoring can be achieved to reduce the background absorption OD below 0.2, which can increase the storage efficiency exceeding 30\%. 
Due to the inhomogeneous broadening of the optical transitions of erbium ions in YSO crystal, memory bandwidth of 500~MHz can be created, which can be integrated with entangled photon pairs generated by spontaneous parametric down-conversion or four-wave-mixing processes to form an efficient light-matter interface \cite{jiang2023quantum}. 
Moreover, the quantum memory devices with two frequency windows can be utilized for coherent processing of time and frequency bin qubits as well as storage of hyperentangled photons with frequency, polarization, time, and number basis. 
In addition, multiple frequency windows can be created to store photons encoded with high-dimensional frequency-bins \cite{lu2023frequency,awschalom2022roadmap}, by doping erbium ions into silica fibers or LiNbO$_\text{3}$, which gives much larger inhomogeneous broadening above 10~GHz \cite{wei2024quantum, zhang2023telecom}.

\section{Conclusion}

In conclusion, we have developed the interleaved pumping scheme to effectively and efficiently prepare atomic frequency comb in a telecom compatible erbium doped crystal. Importantly, we demonstrate efficiency enhancement by more than an order of magnitude compared to previous work in similar experimental conditions. We illustrate this effect with an Er:YSO crystal under a moderate temperature and magnetic field environment from a table-top cryostat and monolithic rare-earth magnets. Moreover, we demonstrate storage of multidimensional qubits in frequency, time and polarization basis with high fidelity.

\begin{acknowledgments}
We wish to acknowledge Ali N Amiri for help with preparation of the millimeter-scale quarter wave plate.

We acknowledge the support from National Science Foundation Award No. 2410054 and U.S. Department of Energy, Office of Science, Office of Advanced Scientific Computing Research, through the Quantum Internet to Accelerate Scientific Discovery Program under Field Work Proposal No. 3ERKJ381.

\end{acknowledgments}

\bibliography{main}% Produces the bibliography via BibTeX.

\end{document}

%arXivadmin\preprin\nolinenumberst{APS/123-QED}

\title{Supplementary Information: Efficient Storage of Multidimensional Telecom Photons in a Solid-State Quantum Memory}

\author{Zongfeng Li}
\author{Yisheng Lei}%
\author{Trevor Kling}%
\affiliation{%
 Department of Electrical and Computer Engineering and Applied Physics Program, Northwestern University, Evanston, IL 60208, USA
}

\author{Mahdi Hosseini}%
\email{Contact author: mh@northwestern.edu}
\affiliation{%
 Department of Electrical and Computer Engineering and Applied Physics Program, Northwestern University, Evanston, IL 60208, USA
}%
\affiliation{%
 Elmore Family School of Electrical and Computer Engineering, Purdue University, West Lafayette, Indiana 47907, USA
}%
\affiliation{%
 Department of Physics and Astronomy, Purdue University, West Lafayette, Indiana 47907, USA
 }%

\date{\today}% It is always \today, today,

\maketitle

\makeatletter
\renewcommand \thesection{S\@arabic\c@section}
\renewcommand\thetable{S\@arabic\c@table}
\renewcommand \thefigure{S\@arabic\c@figure}
\makeatother

\subsection{Experimental details}\label{sec:expDesign} % Contents: Optical Setup Details

\begin{figure*}[!h]
    \includegraphics[width=\columnwidth]{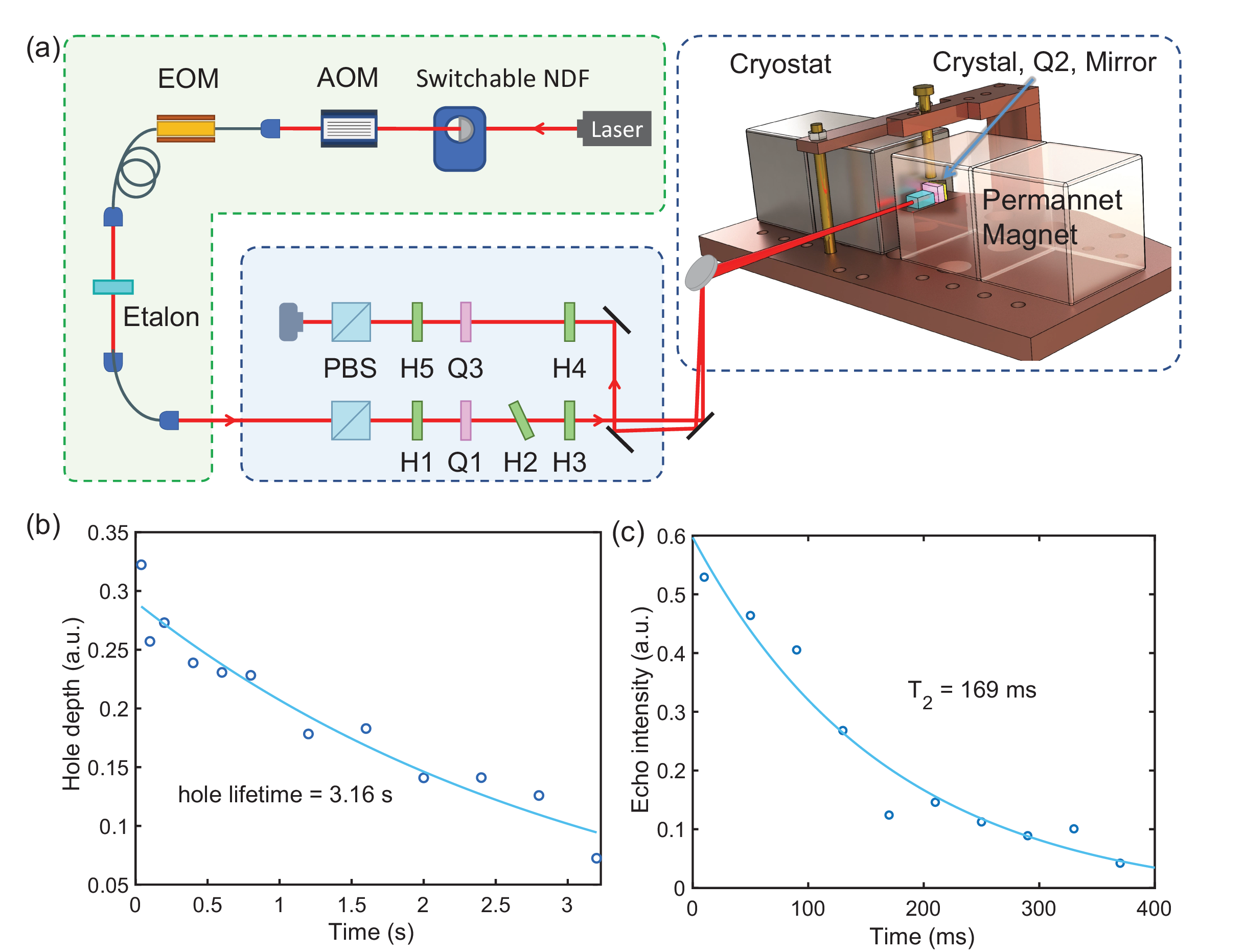}
        \caption{
        (a) Experimental setup for the quantum memory experiment.
        (b) Hole lifetime measured by recording the hole depth at different times after pumping. The solid line is a single-exponential fit.  (c) Two-pulse echo intensity measured at different delays to find the coherence time. }
    \label{fig:si_setup}
\end{figure*}

The optical experiment setup is depicted in Fig. \ref{fig:si_setup} (a). A 1536.3 nm laser is modulated to create pumping and input data sequence by a double-passed acousto-optic modulator (AOM) and a high-speed arbitrary waveform generator. A fiber-coupled electro-optic modulation (EOM) shifts the laser frequency and only the first-order sideband is transmitted through the subsequent etalon.
A rapidly switchable neutral density filters (NDF) is used to attenuate the input pulses into single photon level.
A pair of polarization beam splitters (PBS), half waveplates (HWP, H1/H5) and quarter waveplates (QWP, Q1/Q3) are used to generate/measure a specific polarization state.
Three HWP (H2-4) are used to compensate the phase difference between two polarizations introduced by the elements inside cryostat. The input passes twice through the crystal, with a polarization shift of $\pi/2$ by a QWP (Q2).  
Two pairs of $3/4''$ N52 cube permanent magnets (K\&J Magnetics) with a separation of 4.5 mm provide a relatively uniform magnetic field of 1.1 T.  The entire crystal-magnet system is placed in a "Cryospot 5" cryostat from Photon Spot Inc., and cooled to 0.9 K. The field is measured via either a photodiode for the classical measurements, or an InGaAs/InP single-photon avalanche diode from Micro Photon Devices for the single-photon measurements.
Fig.\ref{fig:si_setup} (b) and (c) show the hole lifetime and the coherence time (measured via two-pulse echo).

\subsection{Optimization of storage efficiency}

To maximize the memory efficiency in this system, we optimized parameters of pumping sequence including the number of pumping loops, pumping power, the duration and bandwidth of a single CHS pulse ($\mathrm{H}_i$) as shown in Fig.~2 in the main text.

\begin{figure*}[!h]
    \includegraphics[width=0.7\columnwidth]{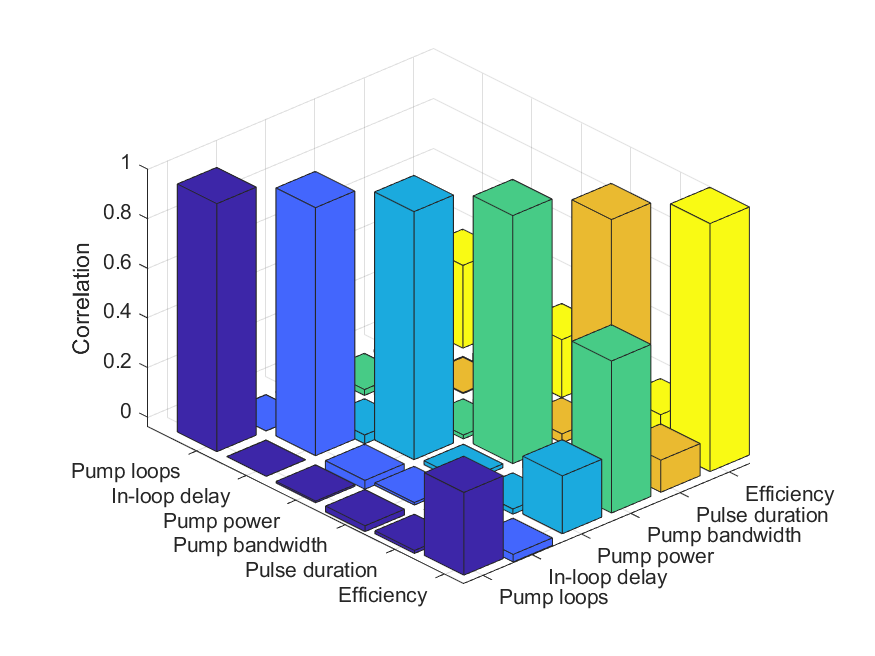}
        \caption{
        Correlation in storage efficiency considering various parameters of initialization and pumping. 
        }
    \label{fig:optimize}
\end{figure*}

Various initialization parameters cannot be independently optimized. 
For this reason we experimentally search for an optimum set of pumping parameters by randomly scanning these parameters through a range of values, and measuring echo efficiency.
Fig.\ref{fig:optimize} illustrates the correlation among parameters from 850 independent experiments conducted in a single optimization iteration.
The first five parameters, being independently sampled, exhibit no correlation.
The in-loop delay parameter exhibits a negligible correlation with efficiency due to the symmetrical relationship between them within the parameter range.
Positive correlations between efficiency and the remaining four parameters suggest that increasing any of these parameters generally improves efficiency.

\begin{figure*}[!h]
    \includegraphics[width=0.7\columnwidth]{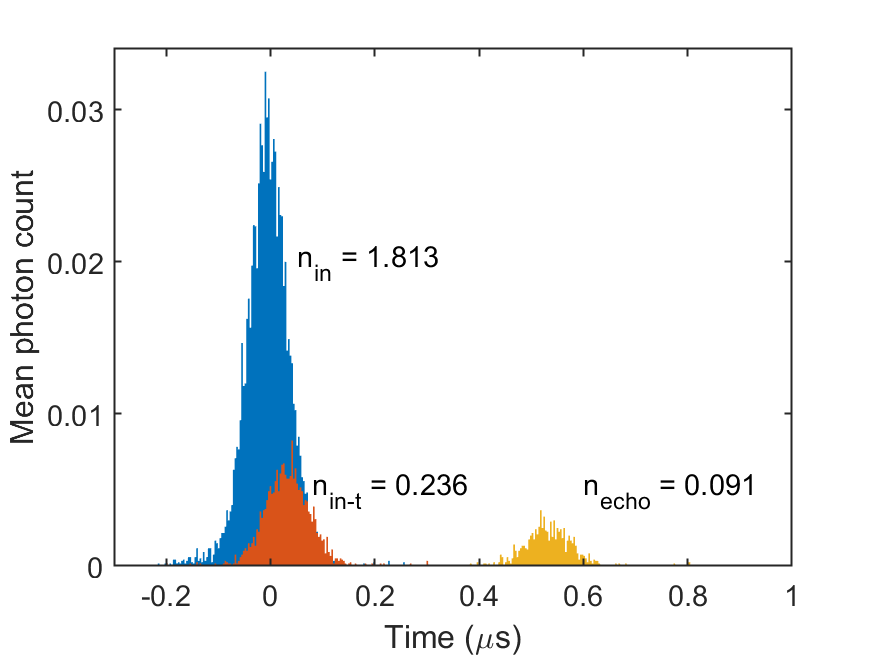}
        \caption{
        Single-photon pulse storage shown for one pulse using a single AFC with storage time of 0.5 $\mathrm{\mu}$s. The mean photon number per pulse is shown for input, leakage, and echo.
        }
    \label{fig:singlePhotonMemory}
\end{figure*}

The best storage efficiency of a classical input pulse is 6.08(0.36)\%, and 5.03\% for the single photon pulse, as shown in Fig. \ref{fig:singlePhotonMemory}.
The original average photon number is calculated using the Poisson distribution to account for detector's saturation.
We also observe the slow light effect in the AFC memory leading to delay of the leakage (unabsorbed) and echo pulses, as seen in Fig.\ref{fig:singlePhotonMemory} \& Fig.3(a). This is due to the spectral dispersion reducing the group velocity of the optical pulses \cite{bonarota2012atomic}.

\subsection{Characterization of frequency and time-bin storage}\label{sec:freq}

\begin{figure*}[!h]
    \includegraphics[width=0.7\columnwidth]{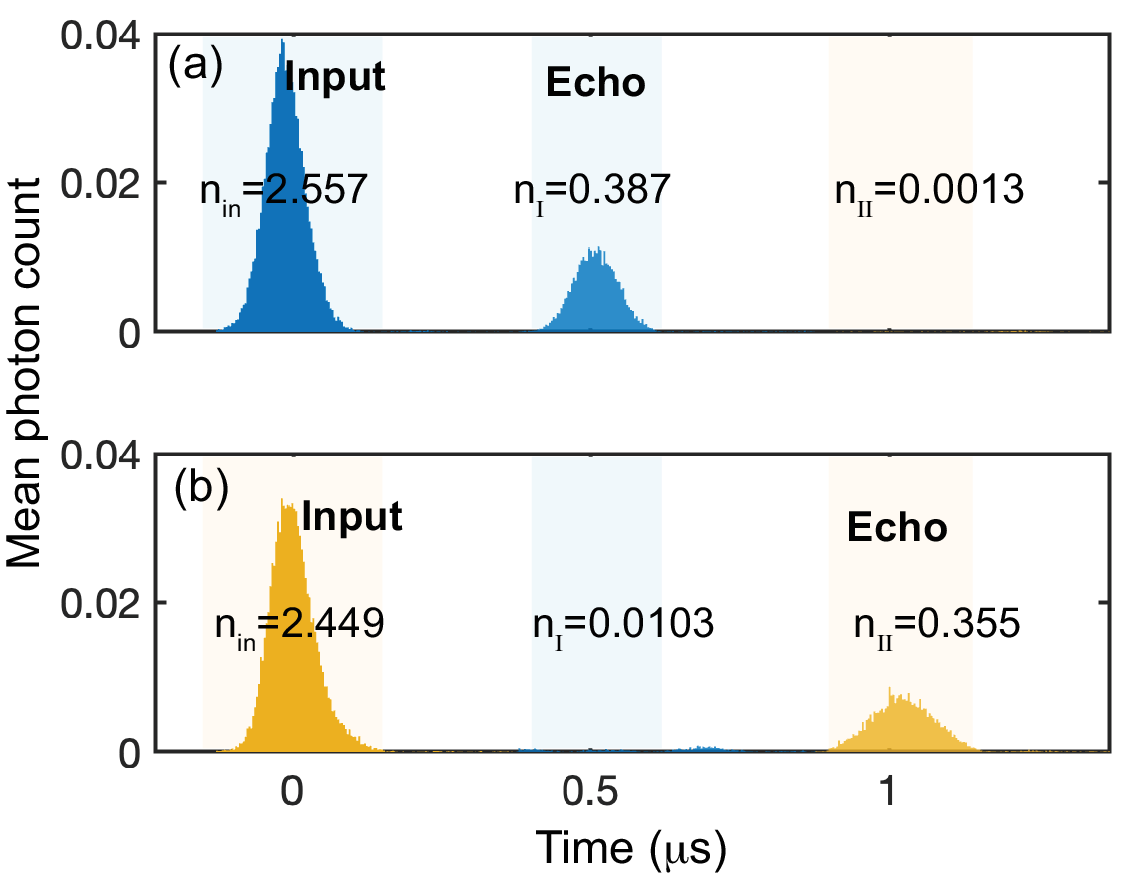}
        \caption{
        Single-pulse storage is demonstrated using a double AFC structure. In (a) and (b), the first and second time bins contain the respective echo signals. Meanwhile, the second and first time bins show negligible photon counts from the AFC window, corresponding to the absence of noise in the respective windows. 
        }
    \label{fig:si_timebin}
\end{figure*}

Multiple AFC windows separated by 50 MHz are used to store distinct frequency qubits. 
As shown in Fig.3(c), time information is encoded to a frequency qubit $(\alpha\ket{0}_f+ \beta\ket{1}_f)\otimes\ket{0}_t \rightarrow \alpha'\ket{0}_f\ket{0}_t + \beta'\ket{1}_f\ket{1}_t$. The fidelity of the frequency qubits can be estimated by setting $\beta=0$, fidelity of $\ket{0}_f$ given by $F_\alpha = \frac{\alpha'}{\alpha'+\beta'}$. Results shown in Fig.\ref{fig:si_timebin} (a) and (b) give a fidelity of 99.7\% and 97.2\%, respectively. The higher counts recorded in the first time bin of Fig.\ref{fig:si_timebin} (b) is due to the inherent property of the AFC, where the comb lines with $2\Delta$ spacing (every-other AFC line) can partially emit an echo at half the expected echo time. This can be resolved by choosing time-bins with temporal spacing different from half memory time.

\subsection{Characterization of polarization storage}\label{sec:pol}

\begin{figure*}[!h]
    \includegraphics[width=\columnwidth]{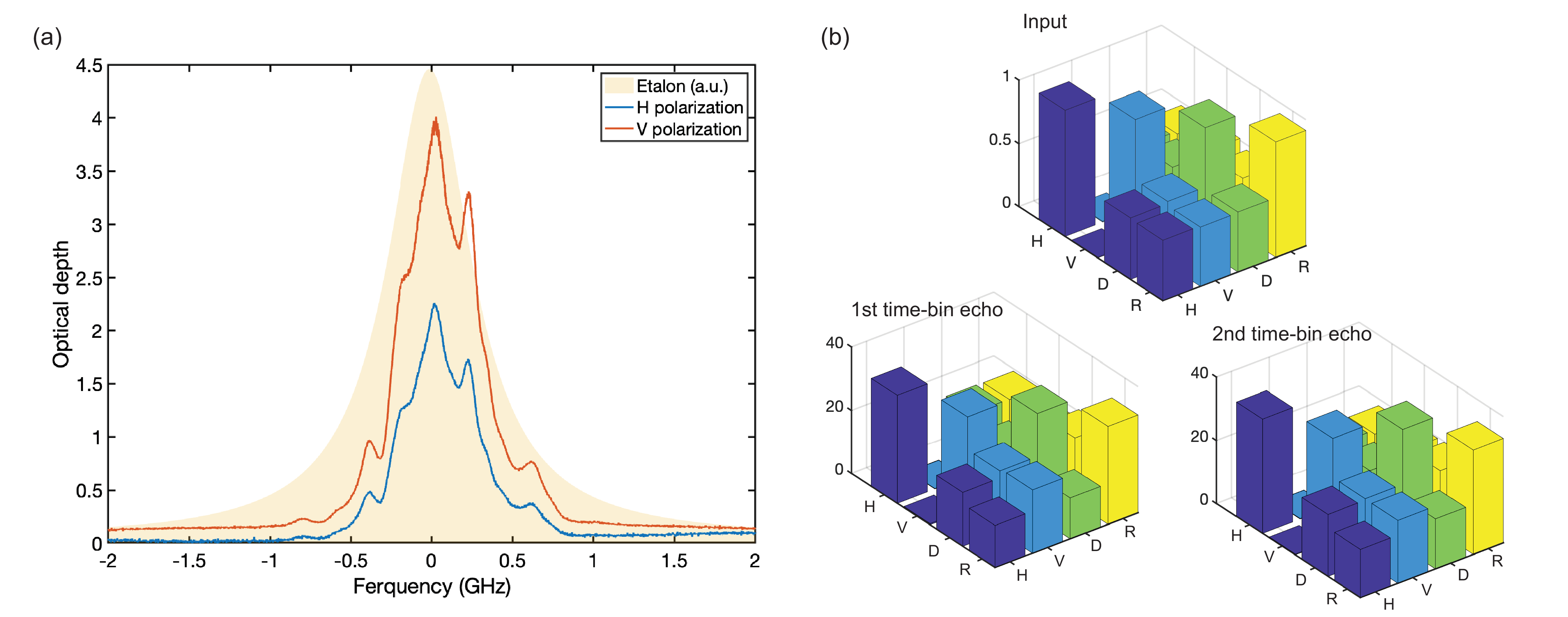}
        \caption{(a) The inhomogeneous linewidth of the crystal for the horizontal and vertical polarizations, within the 1 GHz Etalon bandwidth.  The effective optical depth of the crystal is lower for a horizontally-polarized source than a vertically polarized source.  (b) Classical polarization measurements for the set of basis states employed in the process tomography.  }
    \label{fig:si_polarization}
\end{figure*}

Polarization storage is achieved using a double pass geometry and a quarter waveplate. Crystal absorption for the horizontal and vertical linear polarizations as well as etalon transmission is shown in Fig.\ref{fig:si_polarization} (a). 
The H and V polarization are transferred to D$_1$-D$_2$ bases by HWP H2 and H3.
Even though the OD for different polarizations is different due to the imperfect 45$^\circ$ alignment of the QWP and the birefringence axis of the crystal, the difference in background absorption OD after AFC pumping results in similar storage efficiency. 

We characterize polarization DoF storage by preparing HVDR light polarization and perform classical tomography prior to single-photon experiment. The result of classical memory tomography is shown in Fig.\ref{fig:si_polarization} (b). 

To achieve high polarization storage fidelity, specific wave plates are introduced to correct the polarization errors introduced by the crystal and 
optical elements between the polarization preparation and measurement.
By writing matrix representations of all optical elements acting on a $\{|H\rangle, |V\rangle\}$ polarization basis, the expected polarization state can be tracked via the Jones calculus for standard elements such as mirrors, beam splitters, and wave plates.

\begin{equation*}
    \begin{split}
        \ J_{HWP}(\phi) &= \quad\begin{bmatrix}
            \cos(2\phi) & -\sin(2\phi) \\ \sin(2\phi) & \cos(2\phi)
        \end{bmatrix}, \qquad J_{H0}(\gamma) = \begin{bmatrix}
            1\quad & 0 \\ 0\quad & e^{i\gamma}
        \end{bmatrix}, \\
        \ J_{QWP}(\phi) &= \frac{1}{\sqrt{2}}\begin{bmatrix}
            1-i\cos(2\phi) & -i\sin(2\phi) \\ -i\sin(2\phi) & 1+i\cos(2\phi)
        \end{bmatrix},\\
        \ J_{Arb}(\phi, \gamma) &= \quad\begin{bmatrix}
            \cos(\frac{\gamma}{2})-i\sin(\frac{\gamma}{2})\cos(2\phi) & -i\sin(\frac{\gamma}{2})\sin(2\phi) \\
            -i\sin(\frac{\gamma}{2})\sin(2\phi) & \cos(\frac{\gamma}{2})+i\sin(\frac{\gamma}{2})\cos(2\phi)
        \end{bmatrix}.
    \end{split}
\end{equation*}
 
Where $J_{HWP}$, $J_{QWP}$, $J_{H0}$ and $J_{Arb}$ are the Jones matrices of the half waveplate (HWP), quarter waveplate, 0$^\circ$ tilted HWP, and arbitrary wave plate(AWP). The tilted 0$^\circ$ HWP introduces a phase difference $\gamma$ between $\ket{H}$ and $\ket{V}$, while the AWP can introduce a phase retardation between its fast and slow axes with arbitrary orientation.
The purpose of polarization correction setup is to compensate the an unknown matrix $J_{Arb}$ (induced by the crystal and optical components) into a unit matrix:
\begin{equation*}
    \begin{split}
        \mathbf{I} &= J_{HWP}(\phi_2) \cdot J_{Arb} \cdot J_{HWP}(\phi_1) \cdot J_{H0}(\gamma)  , \\
        \mathbf{I} &\sim J_{Arb} \cdot J_{HWP}(\phi_3) \cdot J_{QWP}(\phi_2) \cdot J_{HWP}(\phi_1).
    \end{split}
\end{equation*}

In the "tilted HWP"-HWP-crystal-HWP configuration (Fig.1(a)), the two HWPs adjacent to the crystal align the {$\ket{H}$, $\ket{V}$} basis with the crystal's fast and slow axes. The 0$^\circ$ tilted HWP pre-compensates for subsequent phase differences.
Additionally, Jones calculus shows that the commonly used HWP-QWP-HWP-crystal configuration is not suitable for the arbitrary compensation.
This is because the normal HWP and the 0$^\circ$ tilted HWP provide rotation operations in the polarization Bloch sphere, allowing for Euler rotations, while the QWP does not contribute to such rotations.

\subsection{Spin-Lattice Coupling} \label{sec:SLC}

\begin{figure*}[!h]
    \includegraphics[width=\columnwidth]{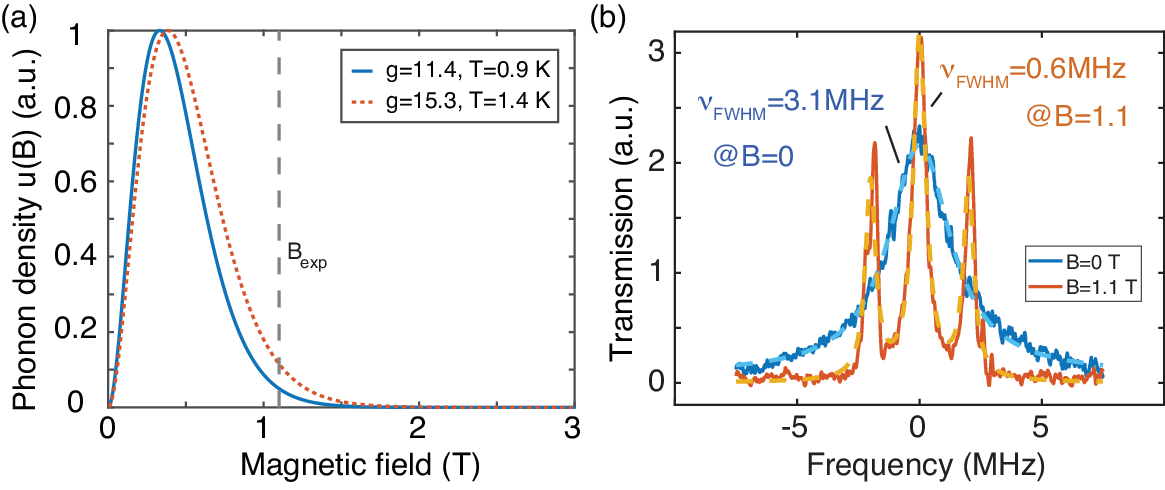}
    \caption{
    (a) Assuming the phonon energy matches the Zeeman splitting, the phonon density distribution in this experiment (blue) is compared to that of site 2 ions in \cite{ranvcic2018coherence} under different magnetic field and temperature (orange dotted).
    (b) The transmission spectra of a hole at 0 and 1.1~T of magnetic field are shown. The hole width ($\nu_{FWHM}$) is found using a Lorentzian fit (dashed line). At high fields, we can resolve two side holes with a spacing proportional to the magnetic field, which we attribute to the differential energy shifts of ground and excited state levels.
    }
    \label{fig:si_phonon}
\end{figure*}

The ground-state lifetime is primarily limited by spin-lattice coupling. Direct, Orbach and Raman phonon transitions induce thermalization. Phonon density and magnetic-field splitting of hyperfine levels can be controlled to reduce these effects. 
It has been shown that around 7~T of magnetic field and 1.4~K temperature, the phonon mediated electronic spin flip can be frozen, significantly enhancing the ground-state lifetime \cite{ranvcic2018coherence}. In the current regime of this experiment, the electron spin is completely frozen, however the phonon density is small enough extending the hyperfine lifetime to more than 3 seconds. 
In addition, in electronic spin frozen regime, spin flip-flops are greatly reduced and the host nuclear spins as well as impurities are partially paramagnetized by strong magnetic field, which greatly narrows the spectral hole linewidth as shown in Fig. \ref{fig:si_phonon}(b) and strongly quenches decoherence for erbium ions \cite{bottger2006spectroscopy, car2019optical}. 
Narrow spectral hole linewidth is crucial for preparing square-like comb lines and small comb spacing $\Delta$, which lead to higher storage efficiency and longer storage time respectively.

Thermal contributions can be quantified by considering the density of phonons with energy matching the splitting of the hyperfine levels of our erbium ions.  The magnetic splitting for a magnetic field along a specified axis can be approximated by $\hbar \omega_{g} =  g\mu_B B$, where $B$ is the magnitude of the applied magnetic field, $g$ is the magnetic g-factor corresponding to the specified axis, and $\mu_B$ is the Bohr magneton.  The density of phonons at a temperature $T$ with a matching frequency can then be approximated via the Planck distribution,
\begin{equation*}
    u(B, T) = \frac{\mu_B^3 g^3 B^3}{\pi^2 \hbar^2 c^3} \frac{1}{e^{\frac{\mu_B g B}{k_B T}}-1}
\end{equation*}
where $k_B$ is the Boltzmann constant.  
The phonon density as a function of the applied field at 0.9~K is shown in Fig. \ref{fig:si_phonon}.
The phonon density is small under the field of 1.1~T and the spin-lattice coupling is partially suppressed.
The spectral hole lifetime of 3~s in this experiment agrees with the phonon density and hole lifetime relation in \cite{ranvcic2018coherence}.

The experimental result of a single spectral hole for zero and 1.1~T of field shown in Fig.\ref{fig:si_phonon} (b) indicates reduced broadening at higher field, where narrower sideholes can be resolved. The side holes are resulted from the differential energy shifts of ground and excited state levels under magnetic field.  By spectral tailoring to create AFC with comb spacing of 2~MHz and 1~MHz we use these side holes to efficiently create the AFC spectrum.

\bibliography{main}% Produces the bibliography via BibTeX.